\documentclass[12pt,preprint]{aastex}

\usepackage{graphicx}
\usepackage{amsbsy}

\begin{document}

\title{Relativistic Particle-In-Cell Simulation Studies of Prompt and Early Afterglows from GRBs}

\author{K.-I. Nishikawa\altaffilmark{1,2},
P. Hardee\altaffilmark{3}, Y. Mizuno\altaffilmark{1,8}, 
M. Medvedev\altaffilmark{4}, \\
B. Zhang\altaffilmark{5}, 
D. H. Hartmann\altaffilmark{6}, and
G. J. Fishman\altaffilmark{7}}

\altaffiltext{1}{National Space Science and Technology Center, 320
Sparkman Drive, VP 62, Huntsville, AL 35805, USA;
ken-ichi.nishikawa-1@nasa.gov} \altaffiltext{2}{Center for Space
Plasma and Aeronomic Research, University of Alabama in Huntsville,
Huntsville, AL 35899, USA}\altaffiltext{3}{Department of Physics
and Astronomy, The University of Alabama, Tuscaloosa, AL 35487, USA}
\altaffiltext{4}{Department of Physics and Astronomy, University of Kansas, KS
66045, USA} 
\altaffiltext{5}{Department of Physics, University of Nevada, Las
Vegas, NV 89154, USA} 
\altaffiltext{6}{Clemson University, Clemson, SC 29634-0978, USA}
\altaffiltext{7}{NASA Marshall Space Flight Center, NSSTC, 
VP62, 320 Sparkman Drive, Huntsville, AL 35805, USA}
\altaffiltext{8}{NASA Postdoctoral Program
Fellow/ NASA Marshall Space Flight Center}

\begin{abstract}
Nonthermal radiation observed from astrophysical systems containing
relativistic jets and shocks, e.g., gamma-ray bursts (GRBs), active
galactic nuclei (AGNs), and microquasars commonly exhibit
power-law emission spectra. Recent PIC simulations of relativistic
electron-ion (or electron-positron) jets injected into a stationary
medium show that particle acceleration occurs within the downstream
jet. In collisionless, relativistic shocks, particle (electron,
positron, and ion) acceleration is due to plasma waves and their
associated instabilities (e.g., the Weibel (filamentation) instability) 
created in the shock region. The simulations show that
the Weibel instability is responsible for generating and amplifying
highly non-uniform, small-scale magnetic fields. These fields
contribute to the electron's transverse deflection behind the jet
head. The resulting ``jitter'' radiation from deflected electrons has 
different properties compared to synchrotron radiation, which assumes a uniform 
magnetic field. Jitter radiation may be important for understanding the
complex time evolution and/or spectra in gamma-ray bursts,
relativistic jets in general, and supernova remnants.
\end{abstract}



\section{Inroduction}

Shocks are believed to be responsible for prompt emission from gamma-ray 
bursts (GRBs) and their afterglows, for variable emission from blazars, 
and for particle acceleration processes in jets from active galactic
nuclei (AGN) and supernova remnants (SNRs). The predominant contribution
to the observed emission spectra is often assumed to be synchrotron- and 
inverse Compton radiation from these accelerated particles (e.g., Piran 1999, 
2000, 2005a; Zhang \& Meszaros 2004; Meszaros 2002, 2006; Lyutikov 2006; 
Zhang 2007). It is assumed that turbulent magnetic fields in the shock region 
lead to Fermi acceleration, producing higher energy particles (e.g.,
Fermi 1949; Blandford \& Eichler 1987). To make progress in understanding 
emission from these object classes, it is essential to place modeling efforts
on a firm physical basis. This requires studies of the microphysics of the 
shock process in a self-consistent manner (Piran 2005b; Waxman 2006).

\subsection{Fermi Acceleration in Test-Particle
Simulations with Turbulent Fields}

Diffusive shock (Fermi) acceleration schemes that have been applied to
non-relativistic shocks have also been applied to particle
acceleration in relativistic shocks (Kirk \& Schneider 1987; Heavens
\& Drury 1988; Bednarz \& Ostrowski 1998; Gallant \& Achterberg 1999;
Kirk et al.\ 2000; Ellison 2001; Achterberg et al.\ 2001; Ellison \&
Double 2002; Vieti 2003; Vladimirov, Ellison \& Bykov 2006; Niemiec \&
Ostrowski 2006; Niemiec, Ostrowski, \& Pohl 2006). Such acceleration
models explicitly, or implicitly, require turbulent conditions
downstream from the shock front and assume a pitch angle diffusion model 
for particle transport near the shock. Particle energies build up
through a cumulative process of particle motion across the shock front
(e.g., Gallant 2002; Niemiec \& Ostrowski 2004). However, highly efficient
acceleration processes required by some observations, e.g., the variable
flux of TeV gamma-rays from Mrk 421 and Mrk 501, are not easy to
reconcile with the diffusive shock acceleration paradigm. For example, 
Bednarek, Kirk, \& Mastichiadis (1996) proposed acceleration by an
electric field to provide a sufficiently fast acceleration process
and to allow for the escape of TeV photons.

There are considerable theoretical problems with the application of
the pitch angle diffusion model to particle transport near
relativistic shocks (Ostrowski \& Bednarz 2002). Diffusive shock
acceleration (DSA) relies on repeated scattering of charged particles
by magnetic irregularities (Alfv\'en waves) to confine the particles
near the shocks. However, in relativistic shocks anisotropies in the
angular distribution of the accelerated particles are large, and the
diffusion approximation for spatial transport breaks down (Achterberg
et al.\ 2001). Despite decades of research, this mechanism is still
not understood from first principles (Waxman 2006). Particle scattering 
in collisionless shocks is due to electromagnetic waves. No present 
analysis self-consistently calculates the generation of these waves, 
scattering of particles, and their acceleration. Most studies 
consider, instead, the evolution of the particle distribution
adopting some {\it Ansatz} for the particle scattering mechanism (e.g.
diffusion in pitch angle), and the ``test particle'' approximation,
where modifications of shock properties due to a population of high 
energy particles is neglected (see, however, Ellison \& Double
2002; Vladimirov, Ellison \& Bykov 2006; Ellison \& Bykov 2008).
Furthermore, the electron spectral index $p$ is calculated, in both
non-relativistic and relativistic cases, with a phenomenological
description of electron scattering based on energy equipartition
($\epsilon_{\rm B}$ and $\epsilon_{\rm e}$) and, therefore, does not
provide a complete, self-consistent description of the process (Piran
2005a,b). In particular, these calculations do not allow one to
determine the fraction of energy carried by electrons (Waxman 2003;
Eichler \& Waxman 2005).

\subsection{Simulation of Particle Acceleration in Relativistic 
Collisionless Shocks and Microscopic Processes}

The problems mentioned in the previous section can be overcome by
detailed, microscopic analyses of energy transfer in collisionless
relativistic outflows (Waxman 2006). Most astrophysical shocks are
collisionless, with energy dissipation dominated by wave-particle
interactions rather than particle-particle collisions (Piran 2005a;
Waxman 2006). In particular, proper study of such relativistic
collisionless shocks in GRB- and AGN jets requires simulating the
microphysics where plasma waves and their associated instabilities
(e.g., the Weibel instability, more precisely, mixed mode two-stream
filamentation instability) simultaneously lead to particle (electron, 
positron, and ion) acceleration and magnetic field generation (Weibel 
1959; Medvedev \& Loeb 1999; Dieckmann et al.\ 2006, references therein).

Three-dimensional relativistic particle-in-cell (RPIC) simulations
have been used to study the microphysical processes in relativistic
shocks. Such PIC simulations show that rapid acceleration takes place
in situ in the downstream jet, rather than by scattering of particles 
back and forth across the shock as in the case of classical Fermi 
acceleration (Silva et al.\ 2003; Frederiksen et al.\ 2003, 2004;
Hededal et al. 2004; Hededal \& Nishikawa 2005; Medvedev et al.\
2005; Nishikawa et al.\ 2003, 2005a,b, 2006a,b; Chang, Spitkovsky \&
Arons 2008; Spitkovsky 2005, 2008). Three independent simulation
studies confirm that relativistic counter-streaming jets do excite
the Weibel instability (Weibel 1959), which generates current filaments 
and associated magnetic fields (Medvedev \& Loeb 1999; Brainerd 2000; 
Pruet et al.\ 2001; Gruzinov 2001; Milosavljevic, Nakar, \& Spitkovsky 
2006; Milosavljevic \& Nakar, 2006a,b), and accelerates electrons (Silva 
et al.\ 2003; Frederiksen et al.\ 2003, 2004; Hededal et al.\ 2004; 
Hededal \& Nishikawa 2005).

\subsection{RPIC Simulations  of Particle
Acceleration and Electromagnetic Field Generation by the Weibel
Instability}


The code used in this study is a modified version of the TRISTAN code,
a relativistic particle-in-cell (RPIC) code (Buneman 1993).
Descriptions of PIC codes can be found in Dawson (1983), Birdsall \&
Langdon (2005), and Hickory \& Eastwood (1988). The RPIC code has been
parallelized using OpenMP on the Columbia computer system at the NASA 
Advanced Supercomputing (NAS) facility and the most recent simulations 
have been performed using the
new parallelized version (Ramirez-Ruiz, Nishikawa \& Hededal
2007). The code has also been parallelized with MPI, and new results
are reported in Niemiec, Pohl, Stroman \& Nishikawa (2008).

The spatial development of a relativistic collisionless shock
involving a moving jet front was investigated in our previous
work (Nishikawa et al.\ 2003, 2005a,b, 2006a,b; Hededal \& Nishikawa
2005). In general, we confirmn the results found in
counter-streaming simulations (Silva et al. 2003; Jaroschek, Lesch,
\& Treumann 2005). Recently, to simulate shock formation, Spitkovsky (2008) 
reflects a relativistically moving cold electron-ion stream from a
conducting wall. This is similar to colliding two streams of identical
plasmas head-on but saves one-half of the computational effort (Chang,
Spitkovsky, \& Arons 2008).
By injecting particle jets into the ISM from one side (left in our simulations) of a 
fixed simulation box, we can study variations in the density of the jet and the 
ambient medium, the density structure, the magnetic field strength and direction, and 
the Lorentz factor, in order to investigate forward and reverse shock development with 
different properties of the fireball ejecta (e.g., Kobayashi et al\. 2007).
In this way we can investigate a relativistically moving system of
precursors, shocks, and contact discontinuities that form in collisions of 
jets with stationary plasmas, with emphasis on radiation signatures from 
the growing instability (e.g., Hoshino 2008; Dieckmann, Shukla, \& Drury
2008). The importance of this kind of simulation, with an injection
scheme without reflection off a conducting wall, is described in the
review paper by Waxman (2006). We have followed this approach to
perform a set of numerical experiments to study the development of 
relativistic collisionless shocks in the context of GRB physics.
Our simulations examine realistic spatial evolution of the resultant 
collisionless shock, including motion of the transition region (the 
contact discontinuity).

\section{Monoenergetic Pair Jet Injected into Electron-Ion Plasmas}

In this section we present a simulation study demonstrating 
how the Weibel instability grows, generates highly structured magnetic
fields, and accelerates particles (Ramirez-Ruiz, Nishikawa, \& Hededal
2007). In particle simulations of relativistic electron-positron jets 
propagating through an unmagnetized electron-positron ambient plasmas, 
the Weibel instability is excited in the downstream region behind the 
jet head and dominates other possible two-stream instabilities (Nishikawa 
et al.\ 2003, 2005a,b, 2006a,b; Hededal \& Nishikawa 2005). This predicted 
result (Brainerd 2000) for relativistic collisionless shocks is different 
from non-relativistic collisionless shocks where other two-stream
instabilities may grow faster than the Weibel instability (Medvedev,
Silva, \& Kamionkowski 2006).

Simulations were performed using an $85 \times 85
\times 640$ grid with a total of 380 million particles (27
particles$/$cell$/$species for the ambient plasma) and an electron
skin depth, $\lambda_{\rm ce} = c/\omega_{\rm pe} = 9.6\Delta$
($\Delta$ is the grid scale), sufficient to study nonlinear spatial
development (Nishikawa et al. 2005a, 2006a). The time step is $t =
0.013/\omega_{\rm pe}$, where $\omega_{\rm pe} = (e^{2}n_{\rm e}/m_{\rm
e})^{1/2}$ is the electron plasma frequency ($n_{\rm e} = n_{\rm b}$, 
where $n_{\rm b}$ is the ambient ``background'' plasma density).  The
simulations described below were performed by the newly parallelized
OpenMP code on Columbia at NASA Advanced Supercomputing (NAS).

\begin{figure*}[h]
\centering
\includegraphics[width=0.5\linewidth]{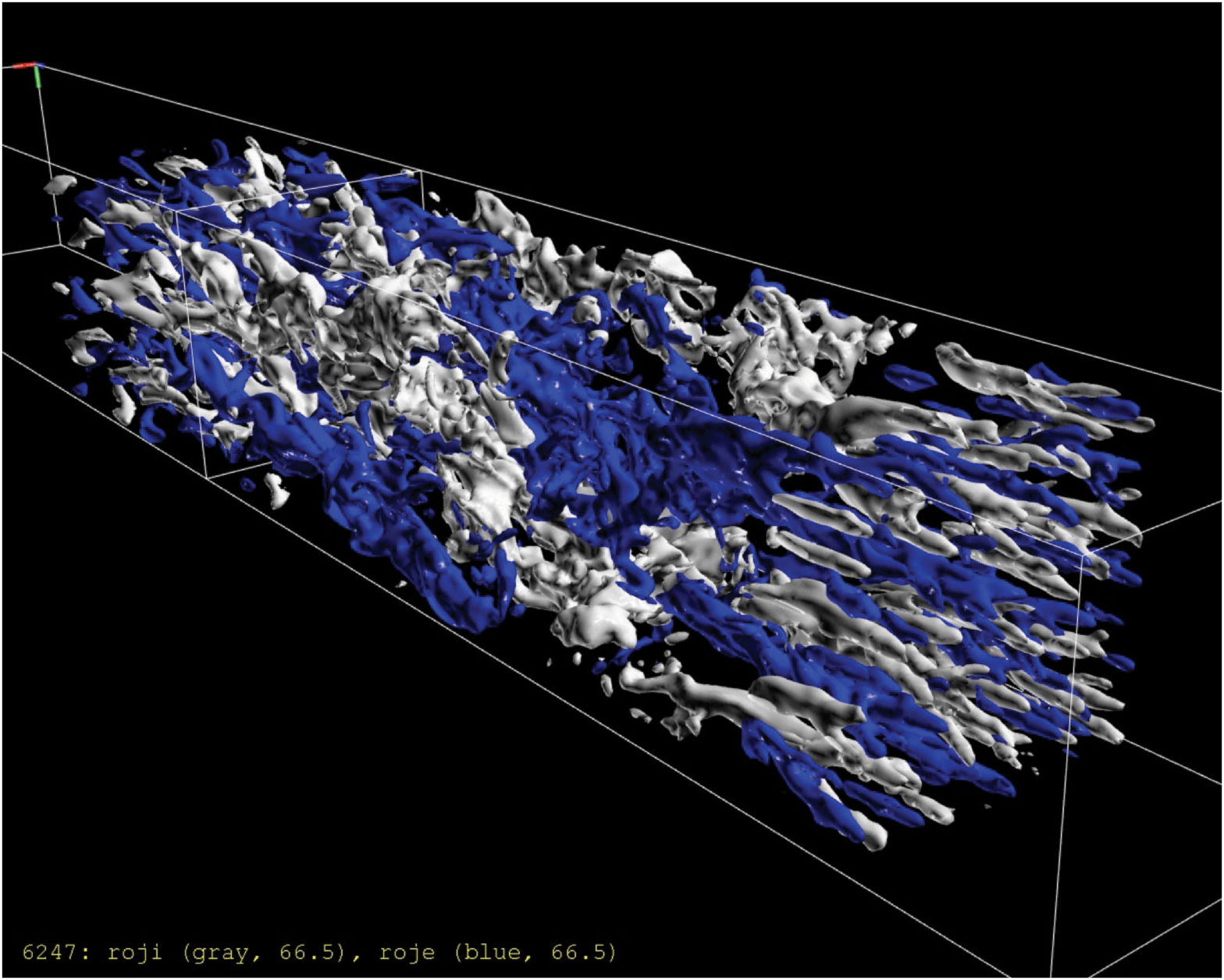}\hspace*{0.2cm}
\includegraphics[width=0.5\linewidth]{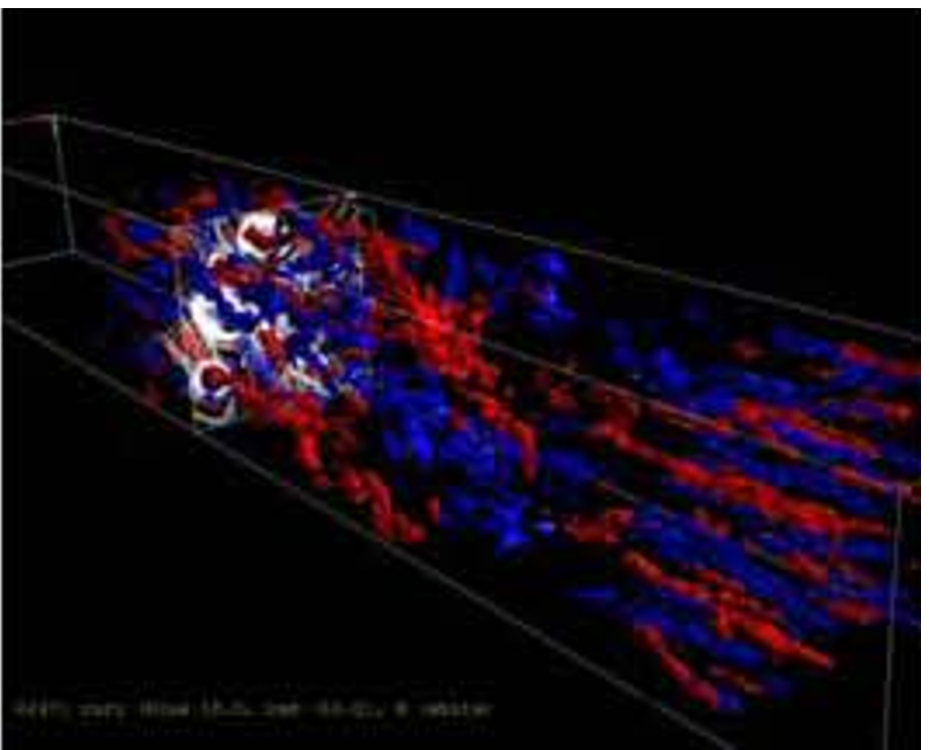}
\caption{Snapshots viewed from the front of the jet at $t =
59.8/\omega_{\rm pe}$, left panel: the isosurface of jet electron
(blue) and positron (gray) density, and right panel: the isosurface of
the $Z$-component of the current density ($J_{\rm z}$ :blue and
$-J_{\rm z}$: red) with the magnetic field lines (white) in the linear
stage for the case of the mono-energetic jet.}
\end{figure*}

We have simulated four different initial pair jet distributions. Here
we present one case, in which a mono-energetic jet ($\gamma V_{\parallel}
= 12.57c$) is injected into an electron-ion ambient plasma, similar to 
published simulations (Nishikawa et al. 2005a, 2006a;
Ramirez-Ruiz, Nishikawa, \& Hededal 2007; Nishikawa et al. 2008).  
For all cases the jet particles are very cold ($0.01 c$ in the rest
frame). The mass ratio of electrons to ions in the ambient plasma is
$m_{\rm i}/m_{\rm e} = 20$. The electron thermal velocity in the
ambient plasma is $v_{\rm th,e} = 0.1c$, where $c$ is the speed of
light. The ion thermal velocity in the ambient plasma is $v_{\rm th,i}
= 0.022c$.

The electron density and currents have a complicated three-dimensional 
structure due to the excitation of the filamentation instability (Ramirez-Ruiz, 
Nishikawa, \& Hededal 2007). Current filaments ($J_{\rm z}$) and their
associated magnetic fields (white curves) produced by the filamentation 
(Weibel) instability form the dominant structures in the relativistic 
collisionless shock shown in Figure 1. In the linear stage, the transverse 
size of these structures is nearly equal to the electron skin depth but 
the longitudinal size (along the jet direction as shown on the left side 
of the left panel) is much larger. Growing smaller current filaments that 
appear first in the linear stage far behind the jet front, eventually merge 
into larger filaments during the nonlinear stage behind the jet front.


\begin{figure}[ht]
\centering
\includegraphics[width=0.83\linewidth]{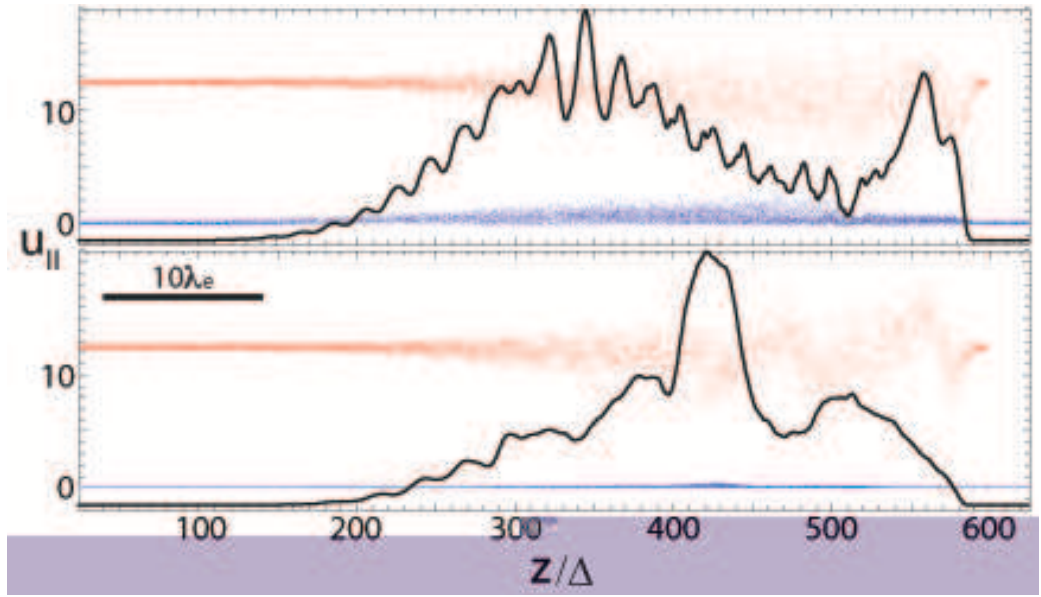}
\caption{\baselineskip 12pt Longitudinal heating and 
acceleration, illustrated by changes in $u_{||}$ for both injected electrons (red) 
and ambient plasma (blue). The bottom (top) panels are for a simulation in 
which the ambient medium is composed of electrons and ions (e$^{\pm}$ pairs). 
Also shown are the average transverse magnetic field amplitude (in arbitrary 
units) in the $X-Y$ plane as a function of $Z/\Delta$ (solid curves).}
\end{figure}

Encountering the medium at rest, the incoming e$^{\pm}$ pairs are rapidly 
deflected by field fluctuations. 
The initial perturbations become non-linear as the deflected e$^{\pm}$ pairs 
collect into 
current channels. The resultant toroidal magnetic fields cause mutual attraction 
between currents, forcing like currents
to approach each other and merge. As a result, the magnetic field grows in 
strength. This continues until the fields grow strong enough to deflect the much 
heavier ions (Frederiksen et al. 2004). 
As illustrated in Fig. 2, the 
ions stay clearly separated in phase space and are only slowly heated. In the
presence of ions, the magnetic field saturates at a higher level, by a factor of 
$(m_{\rm i}/m_{\rm e})^{1/2}=\sqrt{20}\sim 4.5$, albeit on a longer timescale. 
The differences are due to the massive ion bulk momentum constituting a
dominant energy reservoir available for particle heating (Ramirez-Ruiz, 
Nishikawa \& Hededal 2007). We also found that the broadband (not monoenergetic) 
jet sustains the stronger magnetic field over a larger region (see Fig. 8 in 
Ramirez-Ruiz, Nishikawa \& Hededal (2007)). 

\section{Radiation models}
\vspace{-0.1cm} \subsection{The Standard Synchrotron Radiation
Model}

Synchrotron emission is widely assumed to be the most important radiation 
mechanism in the external shock thought to be responsible for the observed 
broad-band afterglows from GRBs (e.g., Zhang
\& Meszaros 2004; Piran 2005a; Zhang 2007, Nakar 2007). Associated
with this picture are three parameterizations that are adopted in almost
all current GRB afterglow models. First, electrons are assumed
to be ``Fermi''-accelerated at the relativistic shocks and to obtain a
power-law distribution in energy, $N(E_{\rm e})dE_{\rm e} \propto E^{-p} 
dE_{\rm e}$, with p $\sim$ 2. This is consistent with numerical simulations 
of shock acceleration (Achterberg et al.\ 2001; Ellisson \& Double 2002; 
Lemoire \& Pelletier 2003). 

Second, the strength and 
geometry of the magnetic fields in the shocked region is unknown, but its 
energy density ($B^{2}/8\pi$) is assumed to be a fraction $\epsilon_{B}$ of 
the internal energy. The values of ``micro-physics'' parameters, such as
$d \log n_{\rm e}/d \log \varepsilon = p$ (the energy distribution of the 
electrons), and the fraction of the energy, $\epsilon_{\rm e}$, carried by electrons, 
are usually obtained by fitting afterglow data (e.g., Panaitescu \& Kumar 2001; 
Yost et al.\ 2003), but are only phenomenological and not based on a
full understanding of the underlying microphysics (Waxman 2006).

The typical observed emission frequency from an electron with
(comoving) energy $\gamma_{\rm e}m_{\rm e}c^{2}$ in a frame with a
bulk Lorentz factor $\Gamma$ is $\nu = \Gamma\gamma_{\rm
e}^{2}(eB/2\pi m_{\rm e}c)$. Three critical frequencies are defined
by three characteristic electron energies. These are $\nu_{\rm m}$
(the injection frequency), $\nu_{\rm c}$ (the cooling frequency),
and $\nu_{\rm M}$ (the maximum synchrotron frequency). In the
afterglow problem, there is one more frequency, $\nu_{\rm a}$, which
is defined by synchrotron self-absorption at lower frequencies
(Meszaros, Rees, \& Wijers 1998; Sari, Piran, \& Narayan 1998; Nakar 
2007; Zhang 2007).

The agreement between the dynamics predicted by the blast wave model
and the direct measurements of the fireball size strongly argue for
the validity of this model's dynamics (e.g., Zhang 2007; Nakar 2007).
The shock wave is most likely collisionless, i.e., mediated by plasma
instabilities (Waxman 2006), and these electromagnetic instabilities
are expected to generate magnetic fields. Afterglow radiation was 
therefore predicted to result from synchrotron emission of shock accelerated 
electrons (Meszaros \& Rees 1997). The observed afterglow spectra are 
indeed remarkably consistent with synchrotron emission of electrons
accelerated to a power-law energy distribution, thus providing support 
to the validity of this "standard afterglow model" (Piran 1999, 2000, 
2005a; Zhang \& Meszaros 2004; Meszaros 2002, 2006; Zhang 2007; Nakar
2007).

In order to determine the luminosity and spectral energy density (SED) 
of synchrotron radiation, the strength of the field ($\epsilon_{\rm B}$)
and the energy distribution of the electrons ($p$) must be determined. 
Due to the lack of a first principles theory of collisionless shocks,  
a purely phenomenological approach to modeling afterglow radiation is 
applied, but one must recognize that emission is then calculated without
a full understanding of the processes responsible for particle acceleration 
and magnetic field generation (Waxman 2006). Despite these shortcomings, it 
is general practise to simply assume that a certain fraction $\epsilon_{\rm B}$ 
of the post-shock thermal energy density is carried by the magnetic field, 
that a fraction $\epsilon_{\rm e}$ is carried by electrons, and that the 
energy distribution of the electrons is a power-law, $d \log n_{\rm e}/d
\log \varepsilon = p$ (above some minimum energy $E_{\rm m}$ which is
determined by $\epsilon_{\rm e}, \epsilon_{\rm B}$ and $p$). In this approach, 
$\epsilon_{\rm B}$, $\epsilon_{\rm e}$, and $p$ are treated as free parameters, 
to be determined by observations. It is important to clarify that
the constraints implied on these parameters by the observations are
independent of any assumptions regarding the nature of the afterglow
shock and the processes responsible for particle acceleration or
magnetic field generation. Any model proposed for the actual shock
micro-physics must be consistent with these phenomenological constraints. 

\subsection{``Jitter'' Radiation from Accelerated
Particles in Turbulent Electromagnetic Fields Generated by the
filamentation (Weibel) Instability}


Since magnetic fields are generated by the current structures produced
in the filamentation (Weibel) instability (Dieckman et al. 2006), it
is possible that ``jitter'' radiation (Medvedev 2000, 2006a,b;
Fleishman 2006a,b; Medvedev et al.\ 2007; Workman et al.\ 2007;
Fleishman \& Toptygin 2007a,b) is an important emission process in
GRBs. It should be noted that synchrotron- and  `jitter'-radiation are 
fundamentally the same physical processes (emission of accelerated
charges in a magnetic field), but the relative importance of the two  
regimes depends on the comparison of the deflection angle and the  
emission angle of the charges (Medvedev 2000). 
Emission via synchrotron- or ``jitter"-radiation from relativistic 
shocks is determined by the magnetic field strength and structure and 
the electron energy distribution behind the shock, which can be computed 
self-consistently with RPIC simulations. The full RPIC  
simulations may actually help to determine whether the emission is more  
synchrotron-like or jitter-like.

The characteristic differences between Syncrotron- and jitter radiation 
are relevant for a more fundamental understanding the complex time evolution 
and/or spectral propertis of GRBs (prompt and afterglows) (Preece et al.\ 1998). 
For example, jitter 
radiation has been proposed as a solution of the puzzle that below their peak 
frequency GRB spectra are sometimes steeper than the ``line of death'' spectral 
index associated with synchrotron emission (Medvedev 2000, 2006a; Fleishman 2006a,b), 
i.e., the observed SED scales as $F_{\nu} \propto \nu^{2/3}$, whereas synchrotron 
SEDs should follow $F_{\nu} \propto \nu^{1/3}$, or even more shallow (e.g., Medvedev 
2006a). Thus, it is crucial to calculate the emerging radiation by tracing electrons 
(positrons) in self-consistently evolved electromagnetic fields. This highly 
complex relativistic radiation-magneto-hydrodynamics-particle-acceleration problem
requires sophisticated tools, such as multi-dimensional, relativistic, PIC methods.

\subsection{New Method of Calculating Synchrotron
and Jitter Emission from Electron Trajectories in Self-consistently
Generated Magnetic Fields}


Consider a particle at position ${\bf{r}_{0}}(t)$ at time $t$ (Fig.\ 3). 
At the same time, we observe the associated electric field from position $\bf{r}$. 
Because of the finite propagation velocity of light, we actually observe the particle 
at an earlier position $\bf{r}_{0}(\rm{t}^{'})$ along its trajectory, labeled with
the retarded time $t^{'} = t - \delta t^{'} = t - \bf{R}(\rm{t}^{'})/c$. Here
$\bf{R}(\rm{t}^{'}) = |\bf{r} - \bf{r}_{0}(\rm{t}^{'})|$ is the distance from the 
charge (at the retarded time $t^{'}$) to the observer's position.

\begin{figure}[ht]
\epsscale{0.37} \vspace*{-0.3cm} \plotone{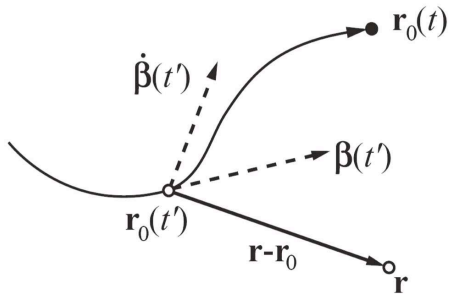}
\vspace*{-0.2cm} \caption{\baselineskip 12pt Definition of the
retardation effect. From an observers point, r, one sees the particle at position 
$\bf{r}_{0}(\rm{t}^{'})$ where it was at retarded time t' (from Figure 2.2 in Hededal 
(2005)).}
\end{figure}

The retarded electric field from a charged particle moving with
instantaneous velocity $\boldsymbol{\beta}$ under acceleration
$\boldsymbol{\dot{\beta}}$ is expressed as (Jackson 1999),

\vspace*{-0.7cm}
\begin{eqnarray}
\bf{E} & = & \frac{q}{4\pi\epsilon_{0}}
\left[\frac{\bf{n}-\boldsymbol{\beta}}{\gamma^{2}
(1-\bf{n}\cdot\boldsymbol{\beta})^{\rm{3}}\rm{R}^{2}}\right]_{\rm ret} +
\frac{q}{4\pi\epsilon_{0}c}
\left[\frac{\bf{n}\times\{(\bf{n}-\boldsymbol{\beta})\times
\boldsymbol{\dot{\beta}}\}}{(1-\bf{n}\cdot\boldsymbol{\beta})^{\rm{3}}\rm{R}}\right]_{\rm ret}
\end{eqnarray}

Here, $\bf{n} \equiv \bf{R}(\rm{t}^{'})/ |\bf{R}(\rm{t}^{'})|$ is a
unit vector that points from the particle's retarded position towards
the observer. The first term on the right hand side, containing the
velocity field, is the Coulomb field from a charge moving without
influence from external forces. The second term is a correction term
that arises when the charge is subject to acceleration. Since the
velocity-dependent field falls off in distance as $R^{-2}$, while the
acceleration-dependent field scales as $R^{-1}$, the latter becomes
dominant when observing the charge at large distances ($R \gg 1$).

The choice of unit vector $\bf{n}$ along the direction of propagation of
the jet (hereafter taken to be the $Z$-axis) corresponds to head-on emission. 
For any other choice of $\bf{n}$ (e.g., $\theta = 1/\gamma$), off-axis emission 
is seen by the observer. The observer's viewing angle is set by the choice of 
$\bf{n}$ ($n_{\rm x}^{2}+n_{\rm y}^{2}+n_{\rm z}^{2} = 1$). 
After some calculation and simplifying assumptions (for detailed
derivation see Hededal 2005) the total energy $W$ radiated per unit
solid angle per unit frequency can be expressed as
\vspace*{-0.3cm}
\begin{eqnarray}
\frac{d^{2}W}{d\Omega d\omega} & = & \frac{\mu_{0} c
q^{2}}{16\pi^{3}} \left|\int^{\infty}_{\infty}\frac{\bf{n}\times
[(\bf{n}-\boldsymbol{\beta})\times \boldsymbol{\dot{\beta}}]}{(1-\boldsymbol{\beta}\cdot
\bf{n})^{2}} e^{i\omega(t^{'} -\bf{n} \cdot \bf{r}_{0}({\rm t}^{'})/{\rm c})}
dt^{'}\right|^{2}
\end{eqnarray}
This equation contains the retarded electric field from a charged particle 
moving with instantaneous velocity $\boldsymbol{\beta}$ under acceleration
$\boldsymbol{\dot{\beta}}$, and only the acceleration field is kept since the
velocity field decreases rapidly as $1/R^{2}$. The distribution over frequencies
of the emitted radiation depends on the particle energy, radius of curvature, and 
acceleration. These quantities are readily obtained from the trajectory of each 
charged particle.

Since the jet plasma has a large velocity $Z$-component in the
simulation frame, the radiation from the particles (electrons and
positrons) is heavily beamed along the $Z$-axis as jitter radiation
(Medvedev 2000, 2006a; Medvedev et al.\ 2007; Workman et al.\ 2007;
Fleishman \& Toptygin 2007a,b).


\section{Radiation from a gyrating electron}

\begin{figure*}[ht]
\centering
\includegraphics[width=0.25\linewidth]{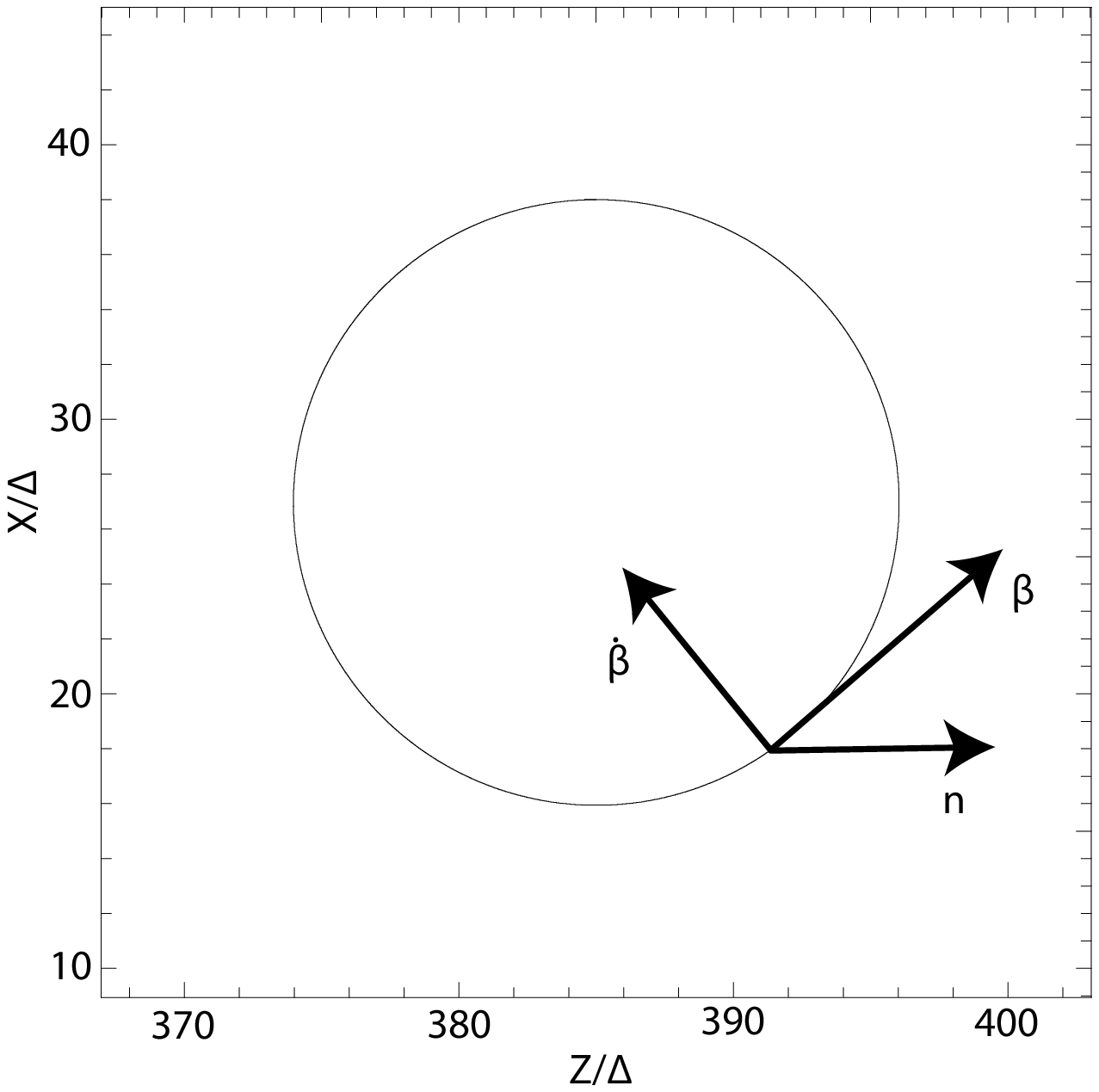}
\includegraphics[width=0.62\linewidth]{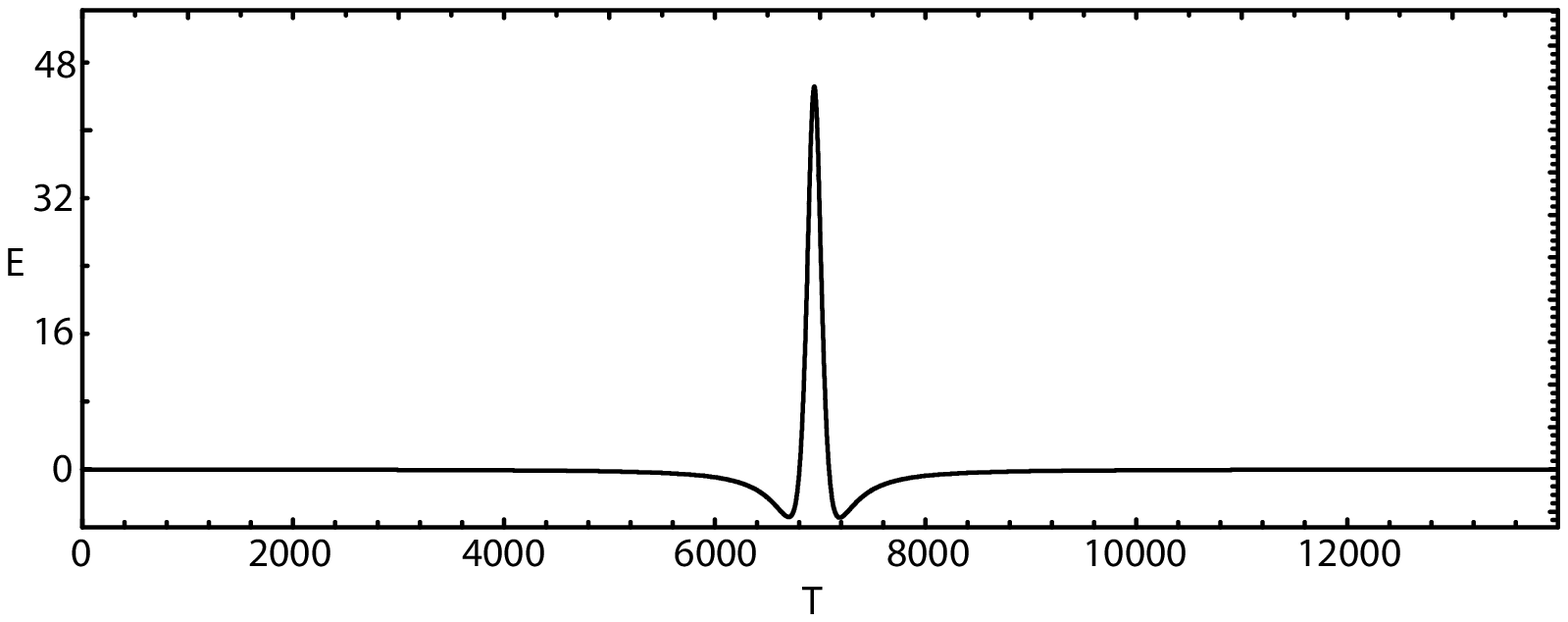}
\caption{The path of a charged particle moving in a homogenous
magnetic field (left panel) (with $\gamma = 15.8$). The particle produces 
a time-dependent, retarded electric field. An observer situated at a large
distance along the {\bf n}-vector sees the retarded electric field
from the gyrating particle (right panel). As a result of relativistic
beaming, the field is seen as pulses peaking when the particle moves
directly towards the observer (Rybicki \& Lightman 1979).}
\centering
\includegraphics[width=0.42\linewidth]{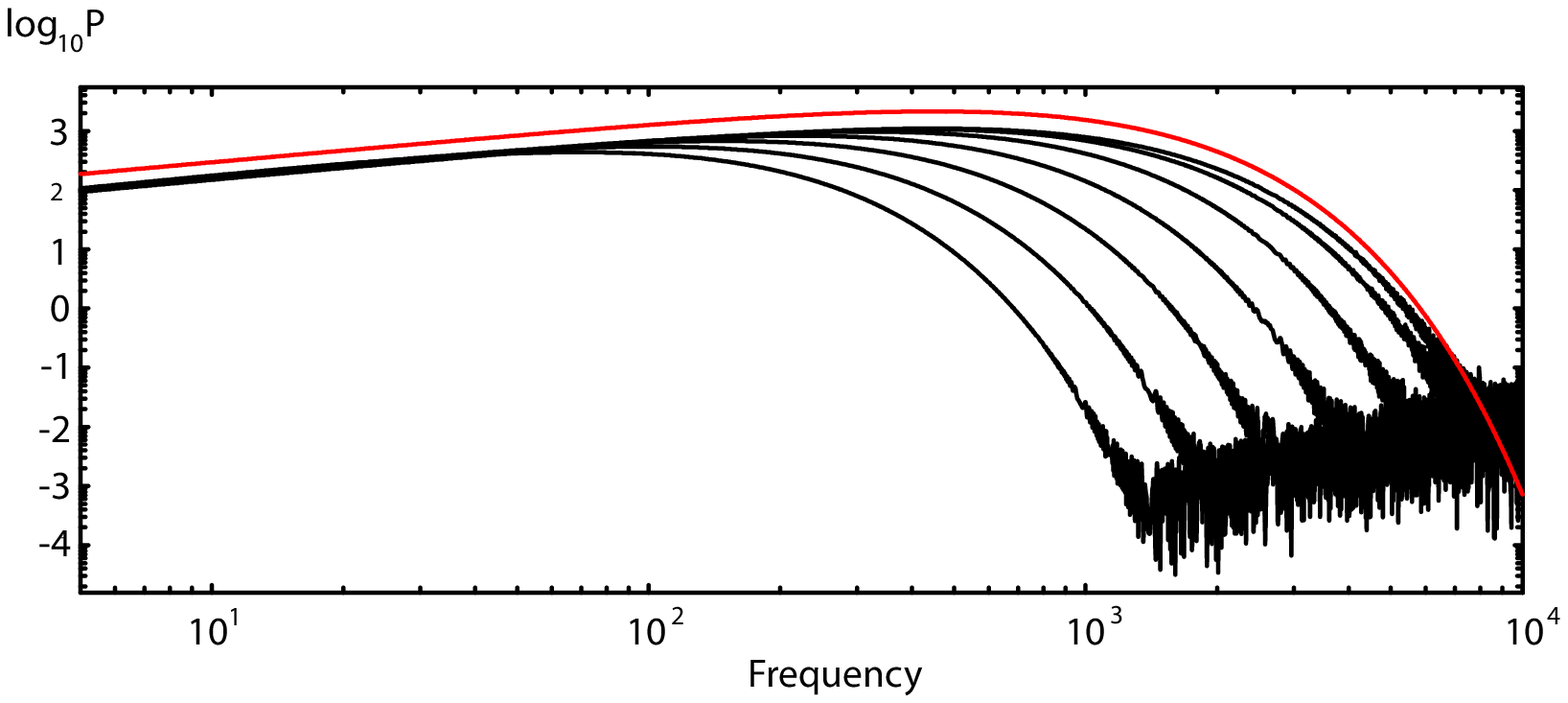}
\vspace*{-0.2cm}
\caption{The observed power spectrum from a single charged particle,
gyrating in a magnetic field at different viewing angles. The viewing
angles are 0$^{\circ}$ (head-on), 1$^{\circ}$, 2$^{\circ}$,
3$^{\circ}$, 4$^{\circ}$, 5$^{\circ}$, and 6$^{\circ}$ ($n_{\rm y}\ne
0$) and the peak frequencies are 448, 408, 318, 222, 148, 98, and 85,
respectively.   At larger angles, frequencies above the Nyquist
frequency are strongly damped. The units on both axes are arbitrary. 
The theoretical synchrotron spectrum  for a viewing angle equal to 
0$^{\circ}$ is plotted for comparison as a  
red curve (multiplied by 2 for clarity).}
\end{figure*}

In the previous section we discussed how to obtain the retarded
electric field from relativistically moving particles (electrons)
observed at large distance. Using eq.\ 2 we calculated the time
evolution of the retarded electric field and the spectrum from a
gyrating electron in a uniform magnetic field to verify the technique
used in this calculation. This calculation agrees with that done by
Hededal (2005). Confirmation of those  results is the first step towards
validation of the implementation of the method in our code. In order
to verify the basic properties of single particle emission (Jackson
1999), we have computed the spectrum for head-on and off angle
observations for two Lorentz factors (15.8 and 40.8).  The angles of off-angle
observations are specified by $n_{\rm y}$ ($n_{\rm y}^{2} + n_{\rm z}^{2} = 1$). 
Here we kept the same gyroradius while increasing the magnetic field strength.
The Nyquist frequency is defined as
$\omega_{\rm N} = 1/2\Delta t$ where $\Delta t$ is the simulation time
step. The frequencies are sampled in a logarithmic scale.

 
\newpage
\begin{figure*}[ht]
\centering
\includegraphics[width=0.25\linewidth]{fig2a.eps}
\includegraphics[width=0.62\linewidth]{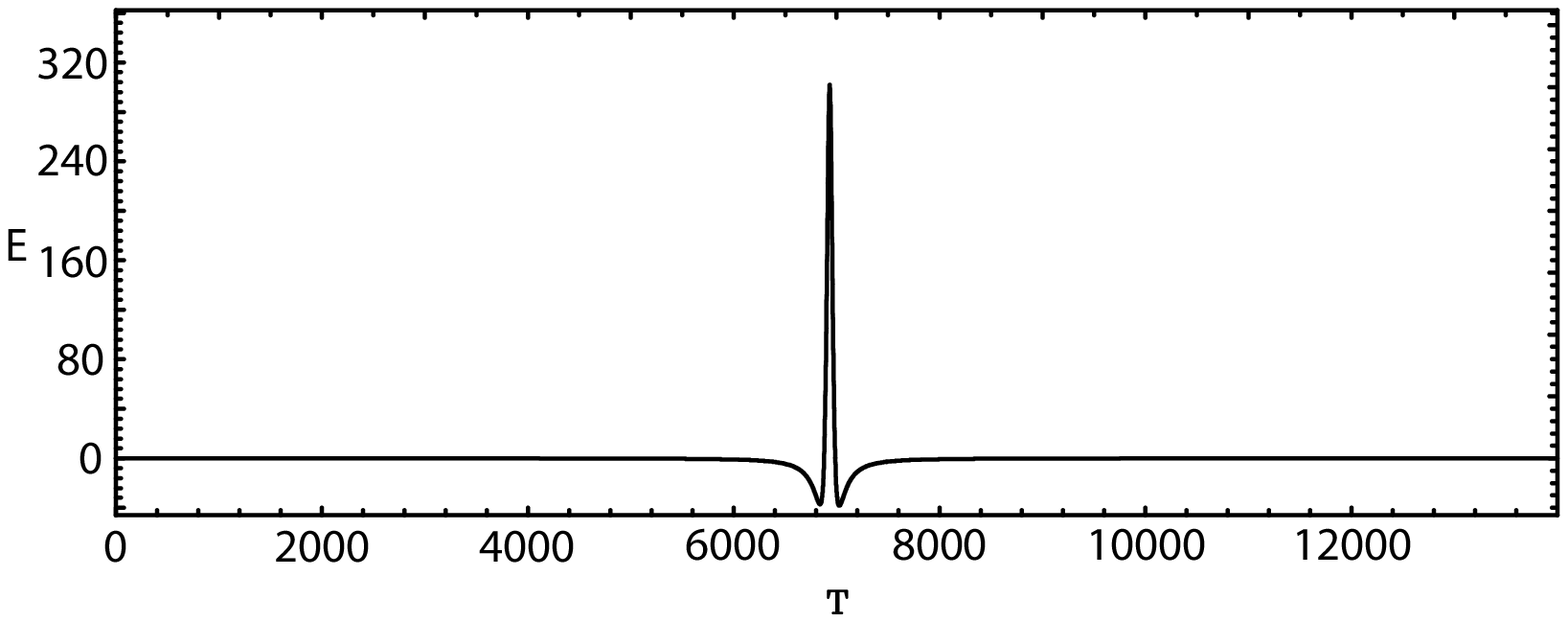}
\caption{The path of a charged particle moving in a homogenous
magnetic field (left panel) ($\gamma = 40.8$). The particle produces a
time dependent electric field. An observer situated at great distance
along the n-vector sees the retarded electric field from the gyrating
particle (right panel). As a result of relativistic beaming, the field
is seen as pulses peaking when the particle moves directly towards the
observer.}
\centering
\includegraphics[width=0.60\linewidth]{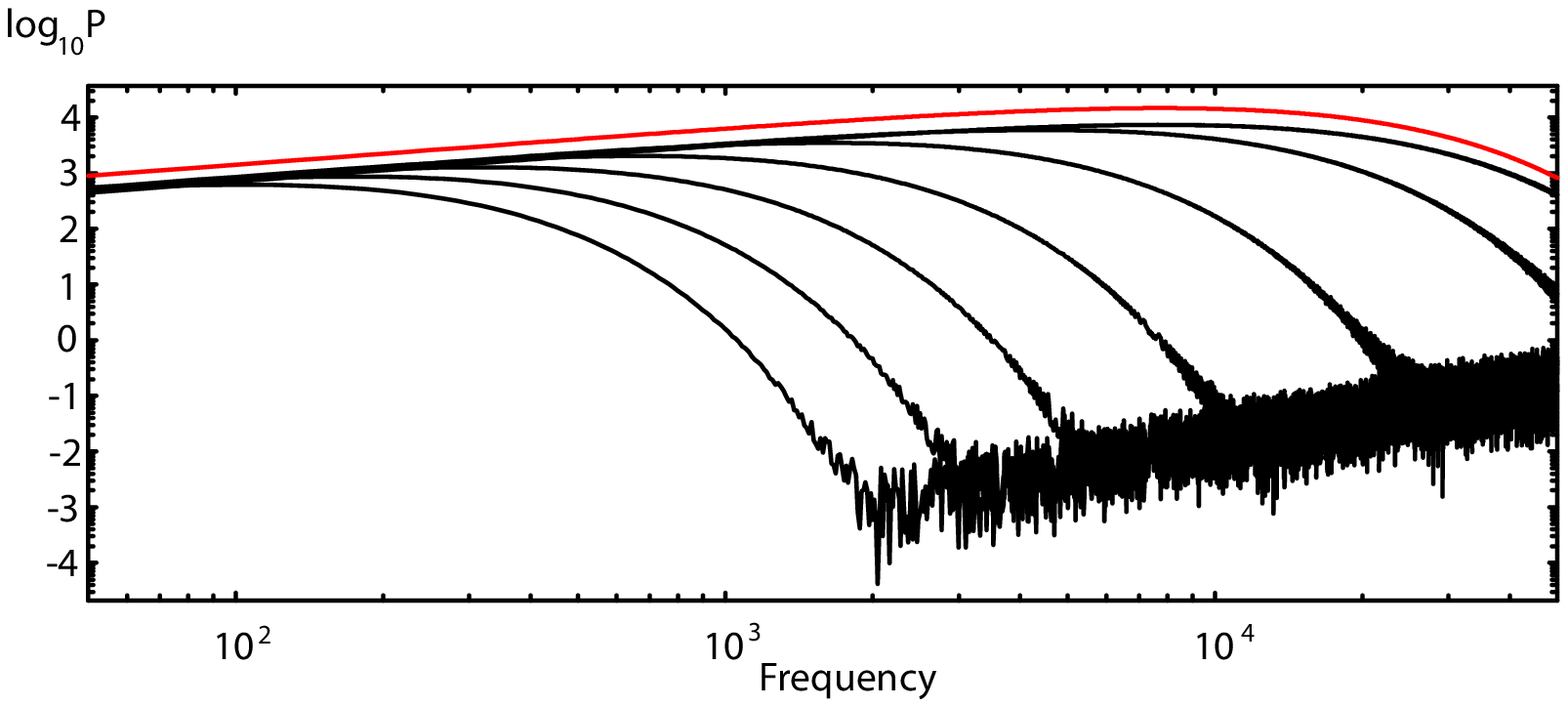}
\caption{The observed power spectrum from a single charged particle,
gyrating in a magnetic field at different viewing angles. The
viewing angles are 0$^{\circ}$, 1$^{\circ}$, 2$^{\circ}$, 3$^{\circ}$,
4$^{\circ}$, 5$^{\circ}$, and 6$^{\circ}$ ($n_{\rm y}\ne 0$) and their
peak frequencies are 7642, 4395, 1648, 666, 316, 166, and 133,
respectively.  The critical frequency $f_{\rm c} =
\frac{3}{2}\gamma^{3}\left(\frac{c}{\rho}\right) = 2309$, where $\rho
= 11.03$.  With larger angles the frequencies above the Nyquist
frequency are strongly damped. The units on both axes are arbitrary. 
The theoretical synchrotron spectrum for a viewing angle equal to 0$^{\circ}$
is plotted for comparison as a red curve (multiplied by 2 for clarity).}  
\end{figure*}

First the case with the lower Lorentz factor ($\gamma = 15.8$) is
calculated. The electron gyrates in the $x-z$-plane with the
uniform magnetic filed ($B_{\rm y}$) and the results are shown in
Figures 4 \& 5. The spectra observed far from the electron at angles
with respect to the $z$ direction are shown in Fig. 5. The critical
frequency $f_{\rm c} =
\frac{3}{2}\gamma^{3}\left(\frac{c}{\rho}\right) = 148$, where $\rho =
11.03$. The higher frequencies ($> f_{\rm c}$) are strongly damped
with increasing angles as $e^{(-f/f_{\rm c})}$, see Jackson (1999) and Melia (2001).
We have very good agreement between the spectrum obtained from the simulation
and the theoretical synchrotron spectrum expectation (red curve) from eq.  3 
(eq. 7.10 (Hededal 2005)).

Synchrotron radiation with the full angular dependency for
the parallel direction is given by (Jackson 1999), 
\begin{eqnarray}
\frac{d^{2}W_{||}}{d\omega d\Omega} &=& \frac{\mu_{0}cq^{2}\omega^{2}}{12\pi}
\left( \frac{r_{\rm L}\theta^{2}_{\beta}\beta^{2}}{c}\right)^{2}\frac{\vert K_{\frac{2}{3}}
(\chi /\sqrt{\cos \theta \beta^{3}})\vert^{2}}{(\cos\theta\beta^{3})^{2}},
\end{eqnarray}

\noindent
where $\theta$ is the angle between {\bf n} and the orbital plane 
$\theta^{2}_{\beta}\equiv 2(1 - \beta\cos\theta)$, $\chi = \omega r_{\rm L}\theta^{3}_{\beta}/3c$ 
and $r_{\rm L}$ the gyro-radius $\gamma mv/(qB)$. For $\beta \rightarrow 1$ and $\theta \rightarrow 0$
this expression converges toward the solution one normally finds in text books
(Jackson 1999; Rybicki \& Lightman 1979; Melia 2001).

For a higher Lorentz factor ($\gamma = 40.8$) several differences are found.  
As expected, the peak value of the retarded electric field is much larger than 
that in the case of a lower Lorentz factor. The width of the spike is narrower,
as shown in Fig.\ 6. The frequencies of the peak value are larger than those 
in the case of lower Lorentz factor, as shown in Fig.\ 7.

These results validate the technique used in our code. It should be noted 
that the method based on the integration of the retarded electric fields calculated
by tracing many electrons described in the previous section can provide a
proper spectrum in turbulent electromagnetic fields. On the other
hand, if the formula for the frequency spectrum of radiation emitted
by a relativistic charged particle in instantaneous circular motion
is used (Jackson 1999; Rybicki \& Lightman 1979), the complex particle
accelerations and trajectories are not properly accounted for and the
jitter radiation spectrum is not properly obtained (for details see 
Hededal 2005).

\section{Discussion}

The procedure used to calculate jitter radiation using the technique
described in the previous section has been implemented in our code.

In order to obtain the spectrum of synchrotron (jitter) emission, we
consider an ensemble of electrons randomly selected in the region
where the filamentation (Weibel) instability has fully developed, and
electrons are accelerated in the generated magnetic fields. We
calculate emission from about 20,000 electrons during the sampling
time, $t_{\rm s} = t_{\rm 2} - t_{\rm 1}$ with Nyquist frequency
$\omega_{\rm N} = 1/2\Delta t$ where $\Delta t$ is the simulation time
step and the frequency resolution $\Delta \omega = 1/t_{\rm
s}$. However, since the emission coordinate frame for each particle is
different, we accumulate radiation at fixed angles in simulation
system coordinates after transforming from the individual particle
emission coordinate frame. This provides an intensity spectrum as a
function of angle relative to the simulation frame $Z$-axis (this can 
be any angle by changing the unit vector ${\bf n}$ in eq.\ (1)). A
hypothetical observer in the ambient medium (viewing the external GRB
shock) views emission along the system $Z$-axis. This computation is
carried out in the reference frame of the ambient medium in the
numerical simulation. For an observer located outside the direction
of bulk motion of the ambient medium, e.g., internal jet shocks in an
ambient medium moving with respect to the observer, an additional
Lorentz transformation would be needed along the line of sight to the
observer.
The spectra obtained from the simulations will be rescaled with a
realistic time scale and relativistic Doppler shift. In electron-ion
jets, the larger mass ratio ($> 100$) will provide enhanced electron
acceleration compared to a mass ratio of 20 used here (Hededal 2005; 
Hededal \& Nordlund 2005; Spitkovsky 2008). 

Emission obtained by the method described above is self-consistent,
and automatically accounts for magnetic field structures on the
small scales responsible for jitter emission. By performing such
calculations for simulations with different parameters, we can then
investigate and compare the quite contrasted regimes of jitter- and
synchrotron-type emission (Medvedev 2000) for prompt and afterglows. 
The feasibility of this approach has been demonstrated and implemented
(Hededal \& Nordlund 2005; Hededal 2005). Thus, we will be able to
address the issue of low frequency GRB spectral index violation of the
synchrotron line of death (Preece et al. 1998; Medvedev 2006a).

Since the emitted radiation is computed during the acceleration step in 
the code we can self-consistently include the effects of radiative losses 
(e.g., Noguchi, Liang, \& Nishimura 2004).
Radiative losses may not affect the global dynamics on our simulation 
timescales, but may be important for particles with the highest Lorentz factors.

\acknowledgments

We have benefited from many useful discussions with E. Ramirez-Ruiz, J. Fredriksen, \AA. 
Nordlund, C. Hededal, and A. J. van der Horst. This work is supported by AST-0506719, 
AST-0506666, NASA-NNG05GK73G and NNX07AJ88G. Simulations were performed at the  
Columbia facility at the NASA
Advanced Supercomputing (NAS) and IBM p690 (Copper) at the National
Center for Supercomputing Applications (NCSA) which is supported by
the NSF. Part of this work was done while K.-I. N. was visiting the
Niels Bohr Institute. He thanks the director of the institution for 
generous hospitality.


\end{document}